\documentclass[a4paper,12pt]{article}\usepackage{graphicx,mathptmx,amsmath,amssymb}
\begin{document}
\title{Kolmogorov complexity as a language}
\author{Alexander Shen\thanks{%
LIF Marseille, CNRS \& Univ.~Aix--Marseille. On leave from IITP, RAS,           
Moscow. Supported in part by NAFIT                                              
ANR-08-EMER-008-01 grant. E-mail: \texttt{alexander.shen@lif.univ-mrs.fr}.
Author is grateful to all the participants of Kolmogorov seminar at Moscow
State University and to his LIF/ESCAPE colleagues. Many of the results covered in this survey were obtained (or at least inspired) by Andrej Muchnik (1958--2007).}}  
\date{}                                                                         
\maketitle  

\begin{abstract}
The notion of Kolmogorov complexity (=the minimal length of a program that generates some object) is often useful as a kind of language that allows us to reformulate some notions and therefore provide new intuition.  In this survey we provide (with minimal comments) many different examples where notions and statements that involve Kolmogorov complexity are compared with their counterparts not involving complexity. 

\end{abstract}

\section{Introduction}

The notion of Kolmogorov complexity is often used as a tool; one may ask, however, whether it is indeed a powerful technique or just a way to present the argument in a more intuitive way (for people accustomed to this notion).

The goal of this paper is to provide a series of examples that support both viewpoints. Each example shows some statements or notions that use complexity, and their counterparts that do not mention complexity. In some cases these two parts are direct translations of each other (and sometimes the equivalence can be proved), in other cases they just have the same underlying intuition but reflect it in different ways. 

Hoping that most readers already know what is Kolmogorov (algorithmic, description) complexity, we still provide a short reminder to fix notation and terminology. The complexity of a bit string $x$ is the minimal length of a program that produces $x$. (The programs are also bit strings; they have no input and may produce binary string as output.) If $D(p)$ is the output of program $p$, the complexity of string $x$ with respect to $D$ is defined as  $K_D(x)=\inf\{ |p|\colon D(p)=x\}$. This definition depends on the choice of programming language (i.e., its interpreter $D$), but we can choose an optimal $D$ that makes $K_D$ minimal (up to $O(1)$ constant). Fixing some optimal $D$, we call $K_D(x)$ the \emph{Kolmogorov complexity} of $x$ and denote it by $K(x)$.  

A technical clarification: there are several different versions of Kolmogorov complexity; if we require the programming language to be self-delimiting or prefix-free (no program is a prefix of another one), we got \emph{prefix} complexity usually denoted by $K(x)$; without this requirement we get \emph{plain} complexity usually denoted by $C(x)$; they are quite close to each other (the difference is $O(\log n)$ for $n$-bit strings and usually can be ignored). 

\emph{Conditional} complexity of a string $x$ given condition $y$ is the minimal length of a program that gets $y$ as input and transforms it into $x$. Again we need to chose an optimal programming language (for programs with input) among all languages. In this way we get \emph{plain conditional complexity} $C(x|y)$; there exists also a prefix version $K(x|y)$.

The value of $C(x)$ can be interpreted as the ``amount of information'' in $x$, measured in bits. The value of $C(x|y)$ measures the amount of information that exists in $x$ but not in $y$, and the difference $I(y:x)=C(x)-C(x|y)$ measures the amount of information in $y$ about $x$. The latter quantity is almost commutative (classical Kolmogorov -- Levin theorem, one of the first results about Kolmogorov complexity) and can be interpreted as ``mutual information'' in $x$ and $y$.

\section{Foundations of probability theory}

\subsection{Random sequences}

One of the motivations for the notion of description complexity was to define randomness:  $n$-bit string is random if it does not have regularities that allow us to describe it much shorter, i.e., if its complexity is close to $n$. For finite strings we do not get a sharp dividing line between random and non-random objects; to get such a line we have to consider infinite sequences. The most popular definition of random infinite sequences was suggested by Per Martin-L\"of. In terms of complexity one can rephrase it as follows: bit sequence $\omega_1\omega_2\ldots$ is random if $K(\omega_1\ldots\omega_n)\ge n-c$ for some $c$ and for all $n$. (This reformulation was suggested by Chaitin; the equivalence was proved by Schnorr and Levin. See more in~\cite{livitanyi,uppsala-notes}.)

Note that in classical probability theory there is no such thing as an individual random object. We say, for example, that randomly generated bit sequence $\omega_1\omega_2\ldots$ satisfies the strong law of large numbers (has limit frequency $\lim (\omega_1+\ldots+\omega_n)/n$ equal to $1/2$) almost surely, but this is just a measure-theoretic statement saying that the set of all $\omega$ with limit frequency $1/2$ has measure $1$. This statement (SLLN) can be proved by using Stirling formula for factorials or Chernoff bound.

Using the notion of Martin-L\"of randomness, we can split this statement into two: (1)~every Martin-L\"of random sequence satisfies SLLN; and (2)~the set of Martin-L\"of random sequences has measure $1$. The second part is a general statement about Martin-L\"of randomness (and is easy to prove). The statement (1) can be proved as follows: if the frequency of ones in a long prefix of $\omega$ deviates significantly from $1/2$, this fact can be used to compress this prefix, e.g., using arithmetic coding or some other technique (Lempel--Ziv compression can be also used), and this is impossible for a random sequence according to the definition.

(In fact this argument is a reformulation of a martingale proof for SLLN.)

Other classical results (e.g., the law of iterated logarithm, ergodic theorem) can be also presented in this way.

\subsection{Sampling random strings}

In the proceeding of this conference S.~Aaronson proves a result that can be considered as a connection between two meanings of the word ``random'' for finite strings. Assume that we bought some device which is marketed as a random number generator. It has some physical source of randomness inside. The advertisement says that, being switched on, this device produces an $n$-bit random string. What could be the exact meaning of this sentence?

There are two ways to understand it. First: the output distribution of this machine is close to the uniform distribution on $n$-bit strings. Second: with high probability the output string is random (=incompressible). The paper of Aaronson establishes some connections between these two interpretations (using some additional machinery).

\section{Counting arguments and existence proofs}

\subsection{A simple example}

Kolmogorov complexity is often used to rephrase counting arguments. We give a simple example (more can be found in~\cite{livitanyi}).

Let us prove by counting that there exists an $n\times n$ bit matrix without $3\log n\times 3\log n$ uniform minors. (We obtain minors by selecting some rows and columns; the minor is \emph{uniform} if all its elements are the same.)

\textbf{Counting}: Let us give an upper bound for the number of matrices with uniform minors. There are at most $n^{3\log n}\times n^{3\log n}$ positions for a minor (we select $3\log n$ rows and $3\log n$ columns).  For each position we have $2$ possibilities for the minor (zeros or ones) and $2^{n^2-(3\log n)^2}$ possibilities for the rest, so the total number of matrices with uniform minors does not exceed
     $$
n^{3\log n} \cdot n^{3\log n} \cdot 2 \cdot 2^{n^2-9\log^2n}=2^{n^2-3\log^2 n +1}< 2^{n^2},
     $$
so there are matrices without uniform minors.

\textbf{Kolmogorov complexity}: Let us prove that incompressible matrix does not have uniform minors. In other words, let us show that matrix with a uniform minor is compressible. Indeed, while listing the elements of such a matrix we do not need to specify all $9\log^2 n$ bits in the uniform minor individually. Instead, it is enough to specify the numbers of the rows of the minor ($3\log n$ numbers; each contains $\log n$ bits) as well as the numbers of columns (this gives together $6\log^2 n$ bits), and to specify the type of the minor ($1$ bit), so we need only $6\log^2 n + 1 \ll 9 \log^2 n$ bits (plus the bits outside the minors, of course).

\subsection{One-tape Turing machines}

One of the first results of computational complexity theory was the proof that some simple operations (checking symmetry or copying) require quadratic time when performed by one-tape Turing machine. This proof becomes very natural if presented in terms of Kolmogorov complexity.

\begin{figure}[h]
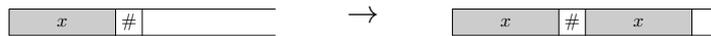

\begin{center}
$\raisebox{-1ex}{\hbox{\includegraphics[scale=0.7]{tape-1.mps}}}
\qquad\to\qquad 
\raisebox{-1ex}{\hbox{\includegraphics[scale=0.7]{tape-2.mps}}}$
\end{center}
\caption{Copying a bit string $x$.}\label{tape-1.mps}
\end{figure}

Assume that initially some string $x$ of length $n$ is written on the tape (followed by the end-marker and empty cells). The task is to copy $x$ just after the marker (Fig.~\ref{tape-1.mps}).

\begin{figure}[h]
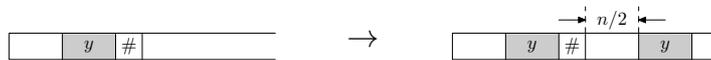

\begin{center}
$\raisebox{-1ex}{\hbox{\includegraphics[scale=0.7]{tape-3.mps}}}
\qquad\to\qquad 
\raisebox{-1ex}{\hbox{\includegraphics[scale=0.7]{tape-4.mps}}}$
\end{center}
\caption{Special case: the first half of $x$ is empty.}\label{tape-3.mps}
\end{figure}

It is convenient to consider a special case of this task when the first half of $x$ is empty (Fig.~\ref{tape-3.mps}) and the second half $y$ is an incompressible string of length $n/2$.

To copy $y$, our machine has to move $n/2$ bits of information across the gap of length $n/2$. Since the amount of information carried by the head of TM is fixed ($\log m$ bits for TM with $m$ states), this requires $\mathrm\Omega(n^2)$ steps (the hidden constant depends on the number of states).

The last statement can be formalized as follows. Fix some borderline inside the gap and install a ``customs office'' that writes down the states of TM when it crosses this border from left to right. This record (together with the office position) is enough to reconstruct $y$  (since the behavior of TM on the right of the border is determined by this record). So the record should be of $\mathrm\Omega(n)$ size. This is true for each of $\mathrm\Omega(n)$ possible positions of the border, and the sum of the record lengths is a lower bound for the number of steps. 

\subsection{Forbidden patterns and everywhere complex sequences}
 
By definition the prefixes of a random sequence have complexity at least $n-O(1)$ where $n$ is the length. Can it be true for all substrings, not only prefixes? No: if it is the case, the sequence at least should be random, and random sequence contains every combination of bits as a substring.

However, Levin noted that the weaker condition $C(x)> \alpha |x| - O(1)$ can be satisfied for all substrings (for any fixed $\alpha<1$).  Such a sequence can be called \emph{$\alpha$-everywhere complex} sequence. Levin suggested a proof of their existence using some properties of Kolmogorov complexity~\cite{dls}.

The combinatorial counterpart of Levin's lemma is the following statement: let $\alpha<1$ be a real number and let $F$ be a set of strings that contains at most $2^{\alpha n}$ strings of length $n$. Then there exists a constant $c$ and a sequence $\omega$ that does not have substrings of length greater than $c$ that belong to $F$.

It can be shown that this combinatorial statement is equivalent to the original formulation (so it can be formally proved used Kolmogorov complexity); however, there are other proofs, and the most natural one uses Lovasz local lemma. (See~\cite{rumyantsev}.) 

\subsection{Gilbert--Varshamov bound and its generalization}

The main problem of coding theory is to find a code with maximal cardinality and given distance. This means that for a given $n$ and given $d$ we want to find some set of $n$-bit strings whose pairwise Hamming distances are at least $d$. The strings are called code words, and we want to have as many of them as possible. There is a lower bound that guarantees the existence of large code, called Gilbert--Varshamov bound.

The condition for Hamming distances guarantees that few (less than $d/2$) bit errors during the transmission do not prevent us from reconstructing the original code word.  This is true only for errors that change some bits; if, say, some bit is deleted and some other bit is inserted in a different place, this kind of error may be irreparable.

It turns out that we can replace Hamming distance by information distance and get almost the same bound for the number of codewords. Consider some family of $n$-bit strings $\{x_1,x_2,\ldots\}$. We say that this family is \emph{$d$-separated}, if $C(x_i|x_j)\ge d$ for $i\ne j$. This means that simple operations of any kind (not only bit changes) cannot transform $x_j$ to $x_i$. Let us show that for every $d$ there exists a $d$-separated family of size $\mathrm\Omega(2^{n-d})$. Indeed, let us choose randomly strings $x_1,\ldots,x_N$ of length $n$. (The value of $N$ will be chosen later.) For given $i$ and $j$ the probability of the event $C(x_i|x_j)<d$ is less than $2^d/2^n$. For given $i$ the probability that $x_i$ is not separated from \emph{some} $x_j$ (in any direction) does not exceed $2N\cdot 2^d/2^n$, so the expected number of $x_i$ that are ``bad'' in this sense is less than $2N^2\cdot 2^d/2^n$. Taking $N=\mathrm\Omega(2^{n-d})$, we can make this expectation less than $N/2$. Then we can take the values of $x_1,\ldots,x_N$ that give less that $N/2$ bad $x_i$ and delete all the bad $x_i$, thus decreasing $N$ at most twice. The decreased $N$ is still $\Omega(2^{n-d})$.

It is easy to see that the Gilbert--Varshamov bound (up to some constant) is a corollary of this simple argument. (See~\cite{chinese} for more applications of this argument.)

\section{Complexity and combinatorial statements}

\subsection{Inequalities for Kolmogorov complexity and their\\ combinatorial meaning}
\label{kolmogorov-levin}

We have already mentioned Kolmogorov--Levin theorem about the symmetry of algorithmic information. In fact, they proved this symmetry as a corollary of the following result:
     $
C(x,y)=C(x) + C(y|x) + O(\log n).
     $
Here $x$ and $y$ are strings of length at most $n$ and $C(x,y)$ is the complexity of some computable encoding of the pair $(x,y)$.

The simple direction of this inequality, $C(x,y)\le C(x)+C(y|x)+O(\log n)$, has equally simple combinatorial meaning. Let $A$ be a finite set of pairs $(x,y)$. Consider the first projection of $A$, i.e., the set $A_X=\{x\colon \exists y\, (x,y)\in A\}$. For each $x$ in $A_X$ we also consider the $x$th section of $A$, i.e., the set $A_x=\{y\colon (x,y)\in A\}$. Now the combinatorial counterpart for the inequality can be formulated as follows: if $\#A_X \le 2^k$ and $\#A_x\le 2^l$ for every $x$, then $\#A \le 2^{k+l}$. (To make the correspondence more clear, we can reformulate the inequality as follows: if $C(x)\le k$ and $C(y|x)\le l$, then $C(x,y)\le k+l+O(\log n)$.)

The more difficult direction, $C(x,y)\ge C(x)+C(y|x)-O(\log n)$, also has a combinatorial counterpart, though more complicated. Let us rewrite this inequality as follows: for every integers $k$ and $l$, if $C(x,y)\le k+l$, then either $C(x)\le k+O(\log n)$ or $C(y|x)\le l+O(\log n)$. It is easy to see that this statement is equivalent to the original one. Now we can easily guess the combinatorial counterpart: if $A$ is a set of pairs that has at most $2^{k+l}$ elements, then one can cover it by two sets $A'$ and $A''$ such that $\#A'_X\le 2^k$ and $\#A''_x\le 2^l$ for every~$x$.

Kolmogorov--Levin theorem implies also the inequality $2C(x,y,z)\le C(x,y)+C(y,z)+C(x,z)$. (Here are below we omit $O(\log n)$ terms, where $n$ is an upper bound of the length for all strings involved.) Indeed, $C(x,y,z)=C(x,y)+C(z|x,y)=C(y,z)+C(x|y,z)$. So the inequality can be rewritten as $C(z|x,y)+C(x|y,z)\le C(x,z)$. It remains to note that $C(x,z)=C(x)+C(z|x)$, that $C(z|x,y)\le C(z|x)$ (more information in the condition makes complexity smaller), and that $C(x|y,z)\le C(x)$ (condition can only help).

The combinatorial counterpart (and the consequence of the inequality about complexities) says that for $A\subset X\times Y\times Z$ we have
     $
(\# A)^2 \le \# A_{X,Y} \cdot \#A_{X,Z} \cdot \#A_{Y,Z},
     $
where $A_{X,Y}$ is the projection of $A$ onto $X\times Y$, i.e., the set of all pairs $(x,y)$ such that $(x,y,z)\in A$ for some $z\in Z$, etc. In geometric terms: if $A$ is a 3-dimensional body, then the square of its volume does not exceed the product of areas of three its projections (onto three orthogonal planes).

\subsection{Common information and graph minors}

We have defined the mutual information in two strings $a,b$ as $I(a:b)=C(b)-C(b|a)$; it is equal (with logarithmic precision) to $C(a)+C(b)-C(a,b)$. The easiest way to construct some strings $a$ and $b$ that have significant amount of mutual information is to take overlapping substrings of a random (incompressible) string; it is easy to see that the mutual information is close to the length (and complexity) of their overlap.

We see that in this case the mutual information is not an abstract quantity, but is materialized as a string (the common part of $a$ and $b$). The natural question arises: is it always the case? i.e., is it possible to find for every pair $a,b$ some string $x$ such that $C(x|a)\approx 0$, $C(x|b)\approx 0$ and $C(x)\approx I(a:b)$?

It turns out that it is not always the case (as found by Andrej Muchnik~\cite{common} in Kolmogorov complexity setting and earlier by G\'acs and K\"orner~\cite{gacs-korner} in Shannon information setting which we do not describe here --- it is not that simple).

The combinatorial counterpart of this question: consider a bipartite graph with (approximately) $2^\alpha$ vertices on the left and $2^\beta$ vertices on the right; assume also that this graph is almost uniform (all vertices in each part have approximately the same degree). Let $2^\gamma$ be the total number of edges. A typical edge connects some vertex $a$ on the left and some vertex $b$ on the right, and corresponds to a pair of complexity $\gamma$ whose first component $a$ has complexity $\alpha$ and second component $b$ has complexity $\beta$, so the ``mutual information'' in this edge is $\delta=\alpha+\beta-\gamma$. The question whether this information can be extracted corresponds to the following combinatorial question: can all (or most) edges of the graph be covered by (approximately) $2^\delta$ minors of size $2^{\alpha-\delta}\times 2^{\beta-\delta}$? (Such a minor connects some $2^{\alpha-\delta}$ vertices on the left with $2^{\beta-\delta}$ vertices on the right.)

For example, consider some finite field $F$ of size $2^n$ and a plane over this field (i.e., two-dimensional vector space). Consider a bipartite graph whose left vertices are points on this plane, right vertices are lines, and edges correspond to incident pairs. We have about $2^{2n}$ vertices is each part, and about $2^{3n}$ edges. This graph does not have $2\times 2$ minors (two different points on a line determine it uniquely). Using this property, one can show that $M\times M$ minor could cover only $O(M\sqrt{M})$ edges. (Assume that $M$ vertices on the left side of such a minor have degrees $d_1,\ldots,d_M$ in the minor. Then for $i$th vertex on the left there are $\mathrm\Omega(d_i^2)$ pairs of neighbor vertices on the right, and all these pairs are different, so $\sum d_i^2 \le O(M^2)$; Cauchy inequality then implies that $\sum d_i \le O(M\sqrt{M})$, and this sum is the number of edges in the minor).

Translating this argument in the complexity language, we get the following statement: for a random pair $(a,b)$ of incident point and line, the complexity of $a$ and $b$ is about $2n$, the complexity of the pair is about $3n$, the mutual information is about $n$, but it is not extractable: there is no string $x$ of complexity $n$ such that $C(x|a)$ and $C(x|b)$ are close to zero. In fact, one can prove that for such a pair $(a,b)$ we have $C(x)\le 2C(x|a)+2C(x|b)+O(\log n)$ for all $x$.

\subsection{Almost uniform sets}

Here is an example of Kolmogorov complexity argument that is difficult to translate to combinatorial language (though one may find a combinatorial proof based on different ideas). Consider the set $A$ of pairs. Let us compare the maximal size of its sections $A_x$ and the average size (that is equal to $\#A/\#A_X$; we use the same notation as in section~\ref{kolmogorov-levin}); the maximal/average ratio will be called $X$-nonuniformity of $A$. We can define $Y$-nonuniformity in the same way.

Claim: \emph{every set $A$ of pairs having cardinality $N$ can be represented as a union of $\mathrm{polylog}(N)$ sets whose $X$- and $Y$-nonuniformity is bounded by $\mathrm{polylog}(N)$}.

Idea of the proof: consider for each pair $(x,y)\in A$ a quintuple of integers 
     $$
p(x,y) = \langle C(x), C(y), C(x|y), C(y|x), C(x,y) \rangle
     $$
where all complexities are taken with additional condition $A$. Each element $(x_0,y_0)$ in $A$ is covered by the set $U(x_0,y_0)$ that consists of all pairs $(x,y)$ for which $p(x,y)\le p(x_0,y_0)$ (coordinate-wise). The number of elements in $U(x_0,y_0)$ is equal to $2^{C(x_0,y_0)}$ up to polynomial in $N$ factors. Indeed, it cannot be greater because $C(x,y)\le C(x_0,y_0)$ for all pairs $(x,y)\in U(x_0,y_0)$. On the other hand, the pair $(x_0,y_0)$ can be described by its ordinal number in the enumeration of all elements of $U(x_0,y_0)$. To construct such an enumeration we need to know only the set $A$ and $p(x_0,y_0)$. The set $A$ is given as a condition, and $p(x_0,y_0)$ has complexity $O(\log N)$. So if the size of $U(x_0,y_0)$ were much less than $2^{C(x_0,y_0)}$, we would get a contradiction.

Similar argument shows that projection $U(x_0,y_0)_X$ has about $2^{C(x_0)}$ elements. Therefore, the average section size is about $2^{C(x_0,y_0)-C(x_0)}$; and the maximal section size does not exceed $C(y_0|x_0)$ since $C(y|x)\le C(y_0|x_0)$ for all $(x,y)\in U(x_0,y_0)$. It remains to note that $C(y_0|x_0)\approx C(x_0,y_0)-C(x_0)$ according to Kolmogorov--Levin theorem, and that there are only polynomially many different sets $U(x,y)$.

Similar argument can be applied to sets of triples, quadruples etc. For a combinatorial proof of this result (in a stronger version) see~\cite{alon}.

\section{Shannon information theory}

\subsection{Shannon coding theorem}

A random variable $\xi$ that has $k$ values with probabilities $p_1,\ldots,p_k$, has \emph{Shannon entropy} $H(\xi)=\sum _i p_i(-\log p_i)$. Shannon coding theorem (in its simplest version) says that if we want to transmit a sequence of $N$ independent values of $\xi$ with small error probability, messages of $NH(\xi)+o(N)$ bits are enough, while messages of $NH(\xi)-o(N)$ bits will lead to error probability close to $1$. 

Kolmogorov complexity reformulation: \emph{with probability close to $1$ the sequence of $N$ independent values of $\xi$ has complexity $NH(\xi)+o(N)$}.

\subsection{Complexity, entropy and group size}

Complexity and entropy are two ways of measuring the amount of information (cf. the title of the Kolmogorov's paper~\cite{kolm65} where he introduced the notion of complexity). So it is not surprising that there are many parallel results. There are even some ``meta-theorems'' that relate both notions. A.~Romashchenko~\cite{romash-ineq} has shown that the linear inequalities that relate complexities of $2^n-1$ tuples made of $n$ strings $a_1,\ldots,a_n$, are the same as for Shannon entropies of tuples made of $n$ random variables. 

In fact, this meta-theorem can be extended to provide combinatorial equivalents for complexity inequalities~\cite{combin}. Moreover, in \cite{group-ineq} it is shown that the same class of inequalities appears when we consider cardinalities of subgroups of some finite group and their intersections!

\subsection{Muchnik's theorem}

Let $a$ and $b$ be two strings. Imagine that somebody knows $b$ and wants to know $a$. Then one needs to send at least $C(a|b)$ bits of information, i.e., the shortest program that transforms $b$ to $a$. However, if we want the message to be not only short, but also simple relative to $a$, the shortest program may not work. Andrej Muchnik~\cite{muchnik-conditional} has shown that it is still possible: \emph{for every two strings $a$ and $b$ of length at most $n$ there exists a string $x$ such that $C(x)\le C(a|b)+O(\log n)$,  $C(a|x,b)=O(\log n)$, and $C(x|a)=O(\log n)$}. This result probably is one of the most fundamental discoveries in Kolmogorov complexity theory of the last decade. It corresponds to Wolf--Slepyan theorem in Shannon information theory; the latter says that for two dependent random variables $\alpha$ and $\beta$ and $N$ independent trials of this pair one can (with high probability) reconstruct $\alpha_1,\ldots,\alpha_N$ from $\beta_1,\ldots,\beta_N$ and some message that is a function of $\alpha_1,\ldots,\alpha_N$ and has bit length close to $NH(\alpha|\beta)$. However, Muchnik and Wolf--Slepyan theorem do not seem to be corollaries of each other (in any direction).

\subsection{Romashchenko's theorem}

Let $\alpha,\beta,\gamma$ be three random variables. The mutual information in $\alpha$ and $\beta$ when $\gamma$ is known is defined as
     $
I(\alpha:\beta|\gamma)=H(\alpha,\gamma)+H(\beta,\gamma)+H(\alpha,\beta,\gamma)-H(\gamma).
    $
It is equal to zero if and only if $\alpha$ and $\beta$ are conditionally independent for every fixed value of $\gamma$.

One can show the following: \emph{If $I(\alpha:\beta|\gamma)=I(\alpha:\gamma|\beta)=I(\beta:\gamma|\alpha)=0$, then one can extract all the common information from $\alpha,\beta,\gamma$ in the following sense\textup: there is a random variable $\chi$ such that $H(\chi|\alpha)=H(\chi|\beta)=H(\chi|\gamma)=0$ and $\alpha,\beta,\gamma$ are independent random variables when $\chi$ is known}. (The latter statement can be written as $I(\alpha:\beta\gamma|\chi)=I(\beta:\alpha\gamma|\chi)=I(\gamma:\alpha\beta|\chi)=0$.)
    
In algebraic terms: if in a $3$-dimensional matrix with non-negative elements all its $2$-dimensional sections have rank $1$, then (after a suitable permutation for each coordinate) it is made of blocks that have tensor rank $1$. (Each block corresponds to some value of $\chi$.)

Romashchenko proved~\cite{romash-triple} a similarly looking result for Kolmogorov complexity: if $a,b,c$ are three strings such that $I(a:b|c)$, $I(b:c|a)$ and $I(a:c|b)$ are close to zero, then there exists $x$ such that $C(x|a)$, $C(x|b)$, $C(x|c)$ are close to zero and strings $a,b,c$ are independent when $x$ is known, i.e., $I(a:bc|x)$, $I(b:ac|x)$ and $I(c:ab|x)$ are close to zero.

This theorem looks like a direct translation of the information theory result above. However, none of these results looks a corollary of the other one, and Romashchenko's proof is a very ingenious and nice argument that has nothing to do with the rather simple proof of the information-theoretic version. 

\section{Computability (recursion) theory}

\subsection{Simple sets}

Long ago Post defined \emph{simple} set as (recursively) enumerable set whose complement is infinite but does not contain an infinite enumerable set (see, e.g., \cite{rogers}, Sect.~8.1). His example of such a set is constructed as follows: let $W_i$ be the $i$th enumerable set; wait until a number $j>2i$ appears in $W_i$ and include first such number $j$ into the enumeration. In this way we enumerate some set $S$ with infinite complement ($S$ may contain at most $n$ integers less than $2n$); on the other hand, $S$ intersects any infinite enumerable set $W_i$, because $W_i$ (being infinite) contains some numbers greater than $2i$.

It is interesting to note that one can construct a natural example of a simple set using Kolmogorov complexity. Let us say that a string $x$ is simple if $C(x)<|x|/2$. The set $S$ of simple strings is enumerable (a short program can be discovered if it exists). The complement of $S$ (the set of ``complex'' strings) is infinite since most $n$-bit strings are incompressible and therefore non-simple. Finally, if there were an infinite enumerable set $x_1,x_2,\ldots$ of non-simple strings, the algorithm ``find the first $x_i$ such that $|x_i|>2n$'' will describe some string of complexity at least $n$ using only $\log n+O(1)$ bits (needed for the binary representation of $n$). 

Similar argument, imitating Berry's paradox, was used by Chaitin to provide a proof for G\"odel incompleteness theorem (see Sect.~\ref{godel}). Note also a (somewhat mystical) coincidence: the word ``simple'' appears in two completely different meanings, and the set of all simple strings turns out to be simple.
 
\subsection{Lower semicomputable random reals}

A real number $\alpha$ is \emph{computable} if there is an algorithm that computes rational approximations to $\alpha$ with any given precision. An old example of E.~Specker shows that a computable series of non-negative rationals can have a finite sum that is not computable. (Let $\{n_1,n_2,\ldots\}$ be a computable enumeration without repetitions of an enumerable undecidable set $K$; then $\sum_i 2^{-n_i}$ is such a series.) Sums of computable series with non-negative rational terms are called \emph{lower semicomputable} reals.
 
The reason why the limit of a computable series is not computable is that the convergence is not effective. One can ask whether one can somehow classify how ineffective the convergence is. There are several approaches. R.~Solovay introduced some reduction on lower semicomputable reals: $\alpha\preceq \beta$ if $\alpha+\gamma=c\beta$ for some lower semicomputable $\gamma$ and some rational $c>0$. Informally, this means that $\alpha$ converges ``better'' than $\beta$ (up to a constant $c$). This partial quasi-ordering has maximal elements called \emph{Solovay complete} reals. It turned out (see~\cite{calude,kucera-slaman}) that Solovay complete reals can be characterized as lower semicomputable reals whose binary expansion is a random sequence.

Another characterization: we may consider the \emph{modulus of convergence}, i.e., a function that for given $n$ gives the first place where the tail of the series becomes less than $2^{-n}$. It turns out that computable series has a random sum if and only if the modulus of convergence grows faster than $BP(n-O(1))$ where $BP(k)$ is the maximal computation time for all terminating $k$-bit self-delimited programs.

\section{Other examples}

\subsection{Constructive proof of Lovasz local lemma}

Lovasz local lemma considers a big (unbounded) number of events that have small probability and are mostly independent. It guarantees that sometimes (with positive probability, may be very small) none of this events happens. We do not give the exact statement but show a typical application: \emph{any CNF made of $k$-literal clauses where each clause has $t=o(2^k)$ neighbors, is satisfiable}. (Neighbors are clauses that have a common variable.)

The original proof by Lovasz (a simple induction proving some lower bound for probabilities) is not constructive in the sense that it does not provide any algorithm to find the satisfying assignment (better than exhaustive search). However, recently Moser discovered that naive algorithm: ``resample clauses that are false until you are done'' converges in polynomial time with high probability, and this can be explained using Kolmogorov complexity. Consider the following procedure (Fig.~\ref{resample}; by \emph{resampling} a clause we mean that all variables in this clause get fresh random values). It is easy to see that this procedure satisfies the specification if terminates (induction).
\begin{figure}[h]
\small
\begin{verbatim}
    {Clause C is false}
    procedure Fix (C: clause)=
      š resample (C);
       for all neighbor clauses C' of C: if C' is false then Fix(C')
    {Clause C is true; all the clauses that were true 
           before the call Fix(C), remain true}  
\end{verbatim}
\caption{Moses' resampling algorithm.}\label{resample}
\end{figure}

The pre- and post-conditions guarantee that we can find a satisfying assignment applying this procedure to all the clauses (assuming the termination). It remains to show that with high probability this procedure terminates in a polynomial time. Imagine that $\texttt{Fix}(X)$ was called for some clause $X$ and this call does not terminate for a long time. We want to get a contradiction. Crucial observation: \emph{at any moment of the computation the sequence of recursive calls made during the execution} (i.e., the ordered list of clauses $C$ for which $\texttt{Fix}(C)$ was called) \emph{together with the current values of all variables determine completely the random bits used for resampling}. (This will allow us to compress the sequence of random bits used for resampling and get a contradiction.) Indeed, we can roll back the computation; note that for every clause in the CNF there is exactly one combination of its variables that makes it false, and our procedure is called only if the clause is false, so we know the values before each resampling.

Now we estimate the number of bits needed to describe the sequence of recursive calls. These calls form a tree. Consider a path that visits all the vertices of this tree (=calls) in the usual way, following the execution process (going from a calling instance to a called one and returning back). Note that called procedure corresponds to one of $t$ neighbors of the calling one, so each step down in the tree can be described by $1+\log t$ bits (we need to say that it is a step down and specify the neighbor). Each step up needs only $1$ bit (since we return to known instance). The number of steps up does not exceed the number of steps down, so we need in total $2+\log t$ bits per call. Since $t=o(2^k)$ by assumption, we can describe the sequence of calls using $k-O(1)$ bits per call which is less than the number of random bits ($k$ per call), so the sequence of calls cannot be long.

\subsection{Berry, G\"odel, Chaitin, Raz}
    \label{godel}
    
Chaitin found (and popularized) a proof of G\"odel incompleteness theorem based on the Berry paradox (``the smallest integer not definable by eight words''). He showed that statements of the form ``$C(x)>k$'' where $x$ is a string and $k$ is a number, can be proved (in some formal theory, e.g., Peano arithmetic) only for bounded values of $k$. Indeed, if it were not the case, we could try all proofs and for every number $n$ effectively find some string $x_n$ which has guaranteed complexity above $n$. Informally, $x_n$ is some string provably not definable by $n$ bits. But it can be defined by $\log n+O(1)$ bits ($\log n$ bits are needed to describe $n$ and $O(1)$ bits describe the algorithm transforming $n$ to $x_n$), so we get a contradiction for large enough $n$. (The difference with the Berry paradox is that $x_n$ is not the minimal string, just the first one in the proofs enumeration ordering.)

Recently Kritchman and Raz found that another paradox, ``Surprise Examination'' (you are told that there will be a surprise examination next week: you realize that it cannot be at Saturday, since then you would know this by Friday evening; so the last possible day is Friday, and if it were at Friday, you would know this by Thursday evening, etc.), can be transformed into a proof of second G\"odel incompleteness theorem; the role of the day of the examination is played by the number of incompressible strings of length $n$. (The argument starts as follows: We can prove that such a string exists; if it were only one string, it can be found by waiting until all other strings turn out to be compressible, so we know there are at least two, etc. In fact you need more delicate argument that uses some properties of Peano arithmetic --- the same properties as in G\"odel's proof.)

\subsection{13th Hilbert problem}

Thirteenth Hilbert problem asked whether some specific function (that gives a root of a degree $7$ polynomial as a function of its coefficients) can be expressed as a composition of continuous functions of one and two real variables. More than fifty years later Kolmogorov and Arnold showed that the answer to this question is positive: any continuous function of several real arguments can be represented as a composition of continuous functions of one variable and addition. (For other classes instead of continuous function this is not the case.) Recently this question was discussed in the framework of circuit complexity~\cite{13H}.

It has also some natural counterpart in Kolmogorov complexity theory. Imagine that three string $a,b,c$ are written on the blackboard. We are allowed to write any string that is simple (has small conditional complexity) relative to any \emph{two} strings on the board, and can do this several times (but not too many: otherwise we can get any string by changing one bit at a time). Which strings could appear if we follow this rule? The necessary condition: strings that appear are simple relative to $(a,b,c)$. It turns out, however, that it is not enough: some strings are simple relative to $(a,b,c)$ but cannot be obtained in this way. This is not difficult to prove (see~\cite{decomposition} for the proof and references); what would be really interesting is to find some specific example, i.e., to give an explicit function with three string arguments such that $f(a,b,c)$ cannot be obtained in the way described starting from random $a$, $b$, and $c$. 

\subsection{Secret sharing}

Imagine some secret (i.e., password) that should be shared among several people in such a way that some (large enough) groups are able to reconstruct the secret while other groups have no information about it. For example, for a secret  $s$ that is an element of the finite field $F$, we can choose a random element $a$ of the same field and make three shares $a$, $a+s$ and $a+2s$ giving them to three participants $X,Y,Z$ respectively. Then each of three participants has no information about the secret $s$, since each share is a uniformly distributed random variable. On the other hand, any two people together can reconstruct the secret. One can say that this secret sharing scheme implements the access structure $\{\{X,Y\},\{X,Z\},\{Y,Z\}\}$ (access structure lists minimal sets of participants that are authorized to know the secret).

Formally, a secret sharing scheme can be defined as a tuple of random variables (one for the secret and one for each participant); the scheme implements some access structure if all groups of participants listed in this structure can uniquely reconstruct the value of the secret, and for all other groups (that do not contain any of the groups listed) their information is independent of the secret. It is easy to see that any access structure can be implemented; the interesting (and open) question is to find how big should be the shares (for a given secret size and a given access structure).

We gave the definition of secret sharing in probability theory framework; however, one can also consider it in Kolmogorov complexity framework. For example, take binary string $s$ as a secret. We may look for three strings $x,y,z$ such that  $C(s|x,y)$, $C(s|y,z)$, and $C(s|x,z)$ are very small (compared to the complexity of the secret itself), as well as the values of $I(x:s)$, $I(y:s)$, and $I(z:s)$. The first requirement means that any two participants know (almost) everything about the secret; the second requirement means each participant alone has (almost) no information about it.

The interesting (and not well studied yet) question is whether these two frameworks are equivalent in some sense (the same access structure can be implemented with the same efficiency); one may also ask whether in Kolmogorov setting the possibility of sharing secret $s$ with given access structure and share sizes depends only on the complexity of $s$. Some partial results were obtained recently by T.~Kaced and A.~Romashchenko (private communication). The use of Kolmogorov complexity in cryptography is discussed in~\cite{antunes}.

\subsection{Quasi-cryptography}

The notion of Kolmogorov complexity can be used to pose some questions that resemble cryptography (though probably are hardly practical). Imagine that some intelligence agency wants to send a message $b$ to its agent. They know that agent has some information $a$. So their message $f$ should be enough to reconstruct $a$ from $b$, i.e., $C(b|a,f)$ should be small. On the other hand, the message $f$ without $a$ should have minimal information about $b$, so the complexity $C(b|f)$ should be maximal.

It is easy to see that $C(b|f)$ cannot exceed $\min(C(a),C(b))$ because both $a$ and $b$ are sufficient to reconstruct $b$ from $f$. Andrej Muchnik proved that indeed this bound is tight, i.e., there is some message $f$ that  reaches it (with logarithmic precision).

Moreover, let us assume that eavesdropper knows some $c$. Then we want to make $C(b|c,f)$ maximal. Muchnik showed that in this case the maximal possible value (for $f$ such that $C(b|a,f)\approx 0$) is $\min(C(a|c),C(b|c))$. He also proved a more difficult result that bounds the size of $f$, at least in the case when $a$ is complex enough. The formal statement of the latter result: \emph{There exists some constant $d$ such that for every strings $a,b,c$ of length at most $N$ such that $C(a|c)\ge C(b|c)+C(b|a)+d\log N$, there exists a string $f$ of length at most $C(b|a)+d\log N$ such that $C(b|a,f)\le d\log N$ and $C(b|c,f)\ge C(b|c)-d\log N$}.

\end{document}